\documentclass[amsmath,amssymb, aps, prb, twocolumn, notitlepage]{revtex4-2}
\usepackage{graphicx}
\usepackage{dcolumn}
\usepackage{bm}
\usepackage[usenames]{color}
\usepackage{slashed}
\usepackage[utf8]{inputenc}
\usepackage{amsfonts}
\usepackage{amsmath}
\usepackage{amssymb}
\usepackage{hyperref}
\usepackage{latexsym}
\usepackage{color}
\usepackage{mhchem}
\usepackage{braket}
\usepackage{caption}

\begin{document}
\title{Density-wave tendency from a topological nodal-line perspective}
\author{Tianlun Zhao, Yi Zhang}
\email{frankzhangyi@gmail.com}
\affiliation{International Center for Quantum Materials, School of Physics, Peking University, Beijing, 100871, China}

\begin{abstract}
The understanding of density waves is a vital component of our insight into electronic quantum matters. Here, we propose an additional mosaic to the existing mechanisms such as Fermi-surface nesting, electron-phonon coupling, and exciton condensation. In particular, we find that certain 2D spin density-wave systems are equivalent to 3D Dirac nodal-line systems in the presence of a magnetic field, whose electronic structure takes the form of Dirac-fermion Landau levels and allows a straightforward analysis of its optimal filling. The subsequent minimum-energy wave vector varies over a continuous range and shows no direct connection to the original Fermi surfaces in 2D. Also, we carry out numerical calculations where the results on model examples support our theory. Our study points out that we have yet to attain a complete understanding of the emergent density wave formalism. 
\end{abstract}

\maketitle

\section{I. Introduction}
Density waves (DW)\cite{RN567, Rn568, RN569, ChO}, including charge density waves (CDW) and spin density waves (SDW), relevant to various physical phenomena in electron and spin quantum matter, have been a fundamental yet controversial problem in condensed matter physics for several decades. In the Peierls theory of CDW, the Fermi surface nesting (FSN) in the 1D chain gives rise to a spatially periodic re-distribution of charge density \cite{XuetaoZhu20150107, FrohlichH1954, M.D.Johannes20080430, Gruner} with a $2\pi/{q}_{n}$ period, commonly accompanied by a distortion of the lattice structure and a metal-insulator transition, where ${q}_{n}=2k_F$ is the nesting vector between the two Fermi points. Though the Peierls transition has successfully described the properties of various quasi-1D DW materials \cite{ChihWeiChen20160704, XuetaoZhu2017}, its extensions to higher dimensions have witnessed many difficulties \cite{XuetaoZhu20150107}. Other mechanisms based upon electron-phonon coupling \cite{M.D.Johannes20060502,PhysRevLett.51.138} and exciton condensation \cite{PhysRev.158.462, PhysRevB.65.235101,Sipos,PhysRevLett.88.226402} offer consistent explanations on the DW origin and physical properties of a series of materials such as $\ce{NbSe_{2}}$, $\ce{TaSe_{2}}$ and $\ce{CeTe_{3}}$ \cite{DEMonciton19750324,C-HDu20000414,C.J.Arguello20140614,MichioNaito19810613, F.Weber20110901} without FSN. Besides, Overhauser's theory pointed out the importance of electron interaction and correlations in the formation of DW in certain materials \cite{PhysRevLett.4.462, PhysRev.128.1437}. 

The existing theories do well in their respective sphere of applications. Some of them are based upon perturbative analysis and become less controlled for stronger coupling \cite{PhysRevLett.51.138, PhysRevB.69.235114}. In addition, most theories cater to preferential DW wave vectors that are discrete and special to the band structures and/or the auxiliary degrees of freedom. Still, the origin of various CDWs, e.g., the 3D DW states in $\ce{M_{3}T_{4}Sn_{13}}$ (e.g. $\ce{Ca_{3}Ir_{4}Sn_{3}}$ and $\ce{Sr_{3}Ir_{4}Sn_{3}}$) \cite{PhysRevLett.109.237008,PhysRevB.86.024522}, the CDW in cuprate materials \cite{PhysRevX.9.021021}, remain controversial to a degree. For instance, the charge modulation in some cuprate materials exhibits a dependence on the spectral gap \cite{PhysRevX.9.021021} with its wave vector spanning a continuous spectrum \cite{Hoffman,sci5584, RN1}. Therefore, our overall understanding of the DW mechanism has not, by far, reached a complete picture yet. 

Here, we propose a novel, independent understanding of DW from a Dirac-fermion Landau levels (LLs) energetics perspective\cite{DzYz}. We note that a class of models with SDW are equivalent to higher-dimensional lattice models in the presence of a magnetic field \cite{Frankicdw2015}. Therefore, to locate the optimal DW wave vector, we can, in turn, look for the magnetic field strength that minimized the energy of the corresponding higher-dimensional system. Such an argument is unrelated to the FSN and remains relevant even when the DW strength is no longer weak in comparison with the bandwidth. Here, we focus on a specific set of two-dimensional (2D) models we find particularly illuminating: these models' counterparts in three dimensions (3D) possess Dirac-fermion nodal lines (NLs), whose electronic structure in a magnetic field possesses a zeroth LL with a large degeneracy and a large gap from the rest of the LLs. Therefore, both the optimal amplitude and direction of the magnetic field and thus the corresponding DW wave vector in the original 2D model depend solely on the geometry of the NL and can be estimated theoretically in a wide parameter region. Also, without changing the Fermi surface, the NL is continuously variable with respect to the DW parameters \cite{PhysRevLett.4.462, PhysRev.128.1437, MC1, MC2}, such as the DW amplitude, so is the optimal DW wave vector \cite{PhysRevB.101.235405, PhysRevB.90.085105, PhysRevB.93.165108, PhysRevB.91.205131}. To solidify our claim, we further illustrate numerical results quantitatively consistent with our theoretical expectations. 

We organize the rest of the paper as follows: In the next section, we introduce the Dirac-fermion LLs perspective for density wave tendencies, including the duality between a 2D DW system and a 3D system with a magnetic field, and the optimal DW wave vector in terms of the energetics on filling the Dirac-fermion LLs. We also summarize the approximations enlisted in our theory and their potential impacts, whose details are also discussed further in Appendix. In section 3, we showcase the emergent DW in a benchmark 2D model and compare the numerical results on the optimal DW wave vectors with our theoretical expectations. Section 4 concludes our discussions with a summary and further implications of our theory and models. In the Appendix, we provide results on several additional models, especially a quasi-1D example showing the compatibility of our theory with the Fermi surface nesting picture. 

\section{II. Theory}


\subsection{Equivalence between a 2D DW system and a 3D system with a magnetic field}
Consider a 2D incommensurate CDW order described by an effective Hamiltonian\cite{PhysRevB.100.045128,Frankicdw2015}:
\begin{equation}
    \hat{H}_{2D,\phi}=\hat{H}_{0}+\sum_{\bm r = (x,y)}V_{q}\cos(\bm{q}\cdot\bm{r}+\phi)\hat{c}^{\dagger}_{r}\hat{c}_{r},
\end{equation}
where $\hat{H}_{0}$ includes the translation-invariant terms, $\hat{c}_{r}$ is the electron annihilation operator at site $\bm{r}$, $V_q$ is the DW amplitude, and $\bm{q}=(q_{x},q_{y})$ is the DW wave vector, is equivalent to a 3D system $\hat{H}_{3D}=\sum_{k_{z}}\hat{H}_{2D,k_{z}}$:
\begin{equation}
\hat{H}_{2D,k_{z}}=\hat{H}_{0,k_z}+\sum_{\bm r = (x,y)}V_{q}\cos(\bm{q}\cdot\bm{r}+k_{z})\hat{c}^{\dagger}_{r,k_z}\hat{c}_{r,k_z},
\end{equation}
with a magnetic field $\bm{B}=(q_{y},-q_{x},0)$ corresponding to the vector potential $\bm{A}=(0,0,\bm{q}\cdot\bm{r})$, and $k_{z}=\phi$ is a good quantum number. We have applied the convention that the lattice spacing is $1$ and $e=\hbar=1$. We aim to determine the optimal $\bm{q}$ with the lowest energy given the rest of the model parameters, in analogy to the search for the Peierls transition in a mean-field treatment of the electron-phonon couplings, etc. A thorough study of a specific model starts from the pristine model with the interactions instead of the density-wave term, which emerges from spontaneous symmetry breaking with minimal total energy. For generality and simplicity, we do not specify an interaction, e.g., electron-electron interaction or electron-boson coupling, focusing on the favorable density-wave scenarios from only the electrons' perspective\cite{YH, SKT}, neglecting or identifying the (energy) contributions from any auxiliary degrees of freedom.

For an incommensurate $\bm{q}$, the bulk physics of $\hat H_{2D,\phi}$ is independent of $\phi=k_{z}$, which is sometimes denoted as a `sliding symmetry.'\cite{PhysRevLett.109.116404,Frankicdw2015} Therefore, $\hat H_{2D,\phi}$ is equivalent to $\hat H_{3D}$ upto a constant factor of $N_{z}$, the number of sites of the 3D system in the $\hat{z}$ direction, and we can analyze the former with the help of the latter (or vice versa). We emphasize that such a mapping seems between a 2D system and a ``$k_z$-slice" of a 3D system for an incommensurate $\bm q$; however, as we analyze the 3D system in the absence of the magnetic field, all $k_z$ values and the entirety of the 3D system matter. The above equivalence relation offers a clear physical picture when the zero-field electronic structure around the Fermi energy takes the form of NLs in the 3D system, which we will discuss next. 

\begin{figure}
\centering
\includegraphics[width=0.9\linewidth]{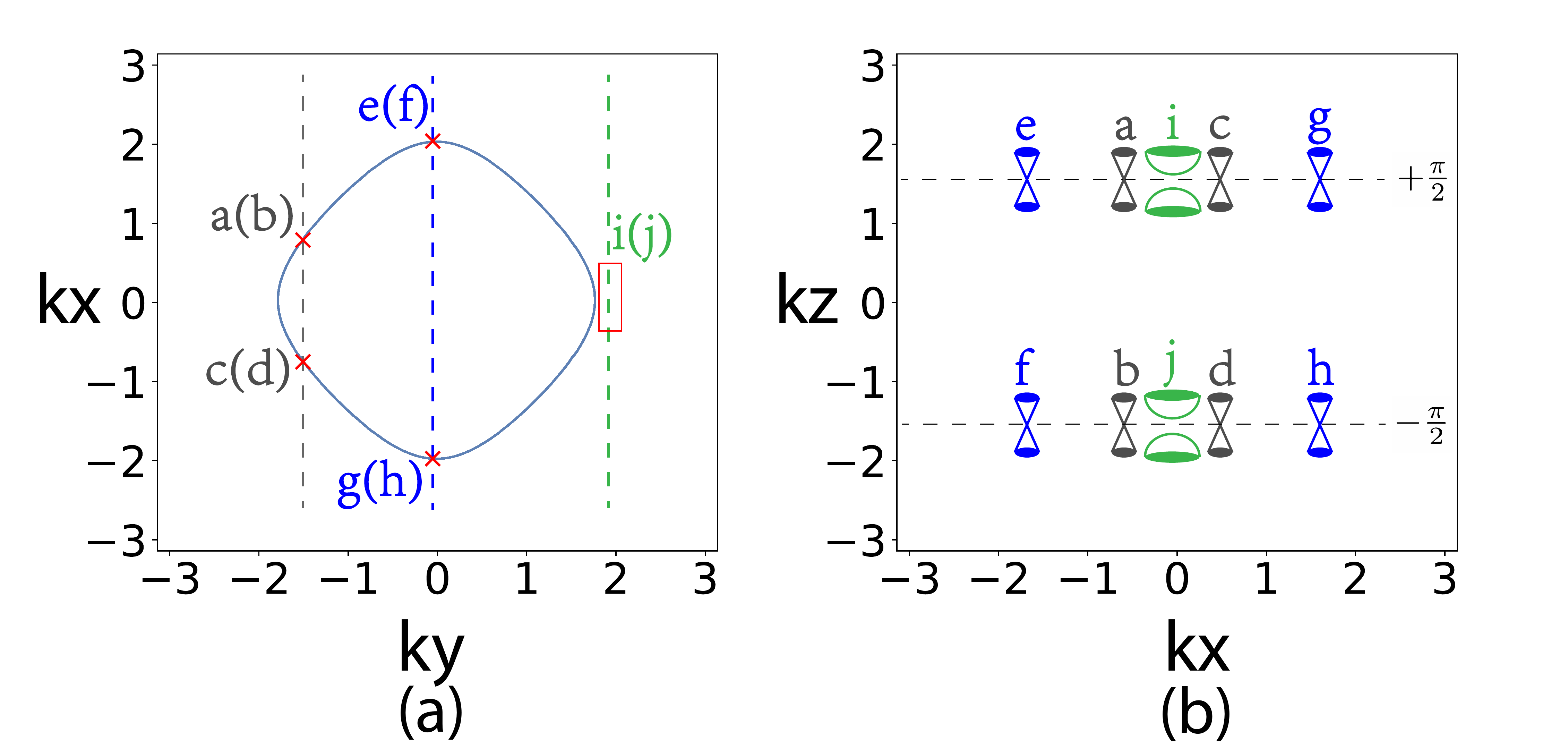}
\caption{(a) The 3D model in Eq. \ref{disper3D} possesses two overlapping nodal lines in the $k_{x}$-$k_{y}$ planes at $k_{z}=\pm\frac{\pi}{2}$ for $h(\bm{k}_{xy})=2t[\cos(k_{x})+\cos(k_{y})]$, $h'(k_z)=2t_{\hat{z}}\cos(k_z)$, $h(k_z)=\epsilon_0=0$, $\epsilon_1=-1$, $t=1$, $t_{\hat{z}}=1$. In a magnetic field along the $\hat{y}$ direction, $k_y$ is a good quantum number that decouples the system into cross-sections intersecting the nodal lines, as the dashed lines. (b) The low-energy massless or massive Dirac fermions in the $k_{x}$-$k_{z}$ plane for different $k_{y}$ values. The colors and labels highlight the correspondence between (a) and (b). } 
\label{nodal}
\end{figure}

\subsection{Dirac-fermion LLs for a NL system}
Let's consider a class of 3D nodal line models in the presence of a magnetic field: 
\begin{equation}
    \begin{split}
        &\hat{H}_{3D} = \hat{H}_{xy}+\hat{H}_{z}+\hat{H}_{\epsilon}\\
        &\hat{H}_{xy}=\sum_{\bm{r},\bm{r}_{xy}}t_{r_{xy}}e^{iA(\bm{r},\bm{r}+\bm{r}_{xy})}\hat{c}^{\dagger}_{r,s}\sigma^{l}_{s,s'}\hat{c}_{r+r_{xy},s'}+h.c.\\
        &\hat{H}_{z}=\sum_{\bm{r}, \bm{r}_{z},p \in \{l,m\}} t_{r_z}e^{iA(\bm{r},\bm{r}+\bm{r}_z)}\hat{c}^{\dagger}_{r,s}\sigma^{p}_{s,s'}\hat{c}_{r+r_{z},s'}+h.c.\\
        &\hat{H}_{\epsilon}=\sum_{\bm{r}}\hat{c}_{r,s}^{\dagger}(\boldsymbol{\epsilon}\cdot\boldsymbol{\sigma})_{s,s'}\hat{c}_{r,s'},
    \end{split}
    \label{SH}
\end{equation}
where $\hat{H}_{xy}$ ($\hat{H}_z$) is the intra-layer (inter-layer) hopping between the $x-y$ planes, $\hat{H}_{\epsilon}$ is the onsite potential and $\boldsymbol{\epsilon}\cdot\boldsymbol{\sigma}=\epsilon_1\sigma^{l}+\epsilon_0\sigma^{m}$, $\sigma^{l}$ and $\sigma^{m}$ are Pauli matrices on the spin or pseudo-spin $s, s'$, $l, m=x,y,z,l\neq m$. $\bm{A}=(0,0,q_{x}x+q_{y}y)$ is the magnetic vector potential of a magnetic field $\bm{B}=(q_{y},-q_{x},0)$. In this gauge, if we only consider the nearest-neighbor hopping in the $\hat{z}$-direction, the Hamiltonian can be diagonalized in the $k_{z}$ basis as:
\begin{equation}
    \begin{split}
        &\hat{H}_{3D}(k_z)=\hat{H}_{xy, k_z}+\hat{H}_{\epsilon,k_z}+\hat{H}_{z,k_z}\\
        &\hat{H}_{z, k_z}=\sum_{\bm{r}',p \in \{l,m\} }V(\bm{r}',k_z)\hat{c}^{\dagger}_{r',k_{z},s}\sigma^{p}_{s,s'}\hat{c}_{r',k_{z},s'}\\
        &V(\bm{r}', k_z)= 2t'_{\hat{z}}\cos (\bm{q}\cdot\bm{r}'-k_{z})- 2t''_{\hat{z}}\sin (\bm{q}\cdot\bm{r}'-k_{z}),\\
    \end{split}
    \label{SH2}
\end{equation}
where $t'_{\hat{z}}$ ($t''_{\hat{z}}$) is the real (imaginary) part of $t_{\hat{z}}$, and $\bm{r}'=(x,y)$.
$\hat{H}_{2D} \propto \hat{H}_{3D}(k_z)$ reflects a 2D SDW system and is our focus in this work. To understand the preferential DW wave vector $\bm{q}$ in such a 2D system, we can analyze the preferential magnetic field amplitude and direction in the equivalent 3D system in Eq. \ref{SH}.

Without the magnetic field, the 3D Hamiltonian $H_{3D}$ can be diagonalized in the $\bm{k}$ space given its fully restored translation symmetry:
\begin{equation}
    \begin{split}
        h_{3D}(\bm{k})&=\left[h(\bm{k}_{xy})+h(k_z)+\epsilon_1\right]\sigma^{l} + \left[h'(k_z)+\epsilon_0\right]\sigma^{m},\\
    \end{split}
    \label{disper3D}   
\end{equation}
where $\bm{k}_{xy}=(k_x, k_y)$, and $h({\bm{k}_{xy}})$ represents the in-plane terms in Eq. \ref{SH}. The Hamiltonian has nodal lines wherever $h(\bm{k}_{xy})+h(k_z)+\epsilon_1=0$ and $h'(k_z)+\epsilon_0=0$, which is illustrated in Fig. \ref{nodal}a. We note that all the nodes on the nodal line are at the same energy in this specific model example, which simplifies our upcoming discussion but can be relaxed to some extent as long as there is little mixing between the zeroth and higher LLs once the magnetic field is present.

In the presence of the magnetic field $\bm{B}$, the momentum $k_\parallel$ parallel to the magnetic field is a good quantum number that labels the different perpendicular cross-sections. Each cross-section may possess pairs of Dirac nodes, as shown in Fig. \ref{nodal}b, which develop into discrete LLs $\epsilon_{n} \propto \pm \sqrt{nB}, n=0,1,2\dots$ in the presence of the magnetic field - the zeroth Landau level exists at the energy of the Dirac nodes, while the rest of the LLs are either above or below with a gap $\propto \sqrt{B}$. Summing over $k_{\parallel}$, we obtain a large zeroth LL degeneracy proportional to the number of Dirac nodes intersected - the NLs' projection along $k_\parallel$, see Fig. \ref{nodes}. The counting works for both strong (large $t_{\hat z}$) and weak (small $t_{\hat z}$) DWs.

\subsection{Magnetic field for optimal filling}
For a system with a fixed electron density $n_e = 1+\delta n_e$, $\delta n_e\ll 1$, the optimal filling is to fill the zeroth LLs and leave all the higher LLs empty. When $|\bm{B}|<|\bm{B}|_{opt}$, the electrons will be forced into the higher LLs, leading to an excitation in an incompressible system and an increase in the systematic energy; when $|\bm{B}|>|\bm{B}|_{opt}$, on the other hand, while the zeroth LLs fully accommodating the electrons above half-filling provide no further energy reduction, the Fermi sea sees an uncompensated energy rise due to the larger magnetic field. Such energy dependence versus the external magnetic field or LL filling constitutes the premise of quantum oscillations \cite{Onsager1952, Lifshitz1956}, e.g., the dHvA effect. Therefore, quantitatively, the electron density above half-filling should match half of the zeroth LL degeneracy:
\begin{equation}
    \begin{split}
        (n_e-1)L_{\parallel}S_{\perp}&=\frac{\int n_{D}(k_{\parallel}) d k_{\parallel}}{2\pi/L_{\parallel}}\frac{|\bm{q}_{opt}|S_{\perp}}{2\pi}\frac{1}{2}\\
        \Rightarrow n_e-1&=\frac{\bar{n}_{D}}{2}\frac{|\bm{q}_{opt}|}{2\pi}, \\    
    \end{split}
    \label{analysis}         
\end{equation}
where $L_{\parallel}$ ($S_{\perp}$) is the length (area) of the system parallel (perpendicular) to the magnetic field, $|\bm{B}|/2\pi=|\bm{q}|/2\pi$ is the LL degeneracy over unit space, $n_{D}(k_{\parallel})$ denotes the number of Dirac nodes in the cross-section at $k_{\parallel}$, and $\bar{n}_{D}=\int n_{D}(k_{\parallel}) d k_{\parallel}/2\pi$ is its average over all $k_{\parallel}$. With $\bar{n}_{D}$, the expression for optimal $|\bm{q}_{opt}|$ resembles the one-dimensional case in Ref. \cite{DFLL2021}, yet $\bar{n}_{D}$ is a continuous variable instead of an integer just like $n_{D}(k_{\parallel})$.

\begin{figure}
\centering
\includegraphics[width = 0.9\linewidth]{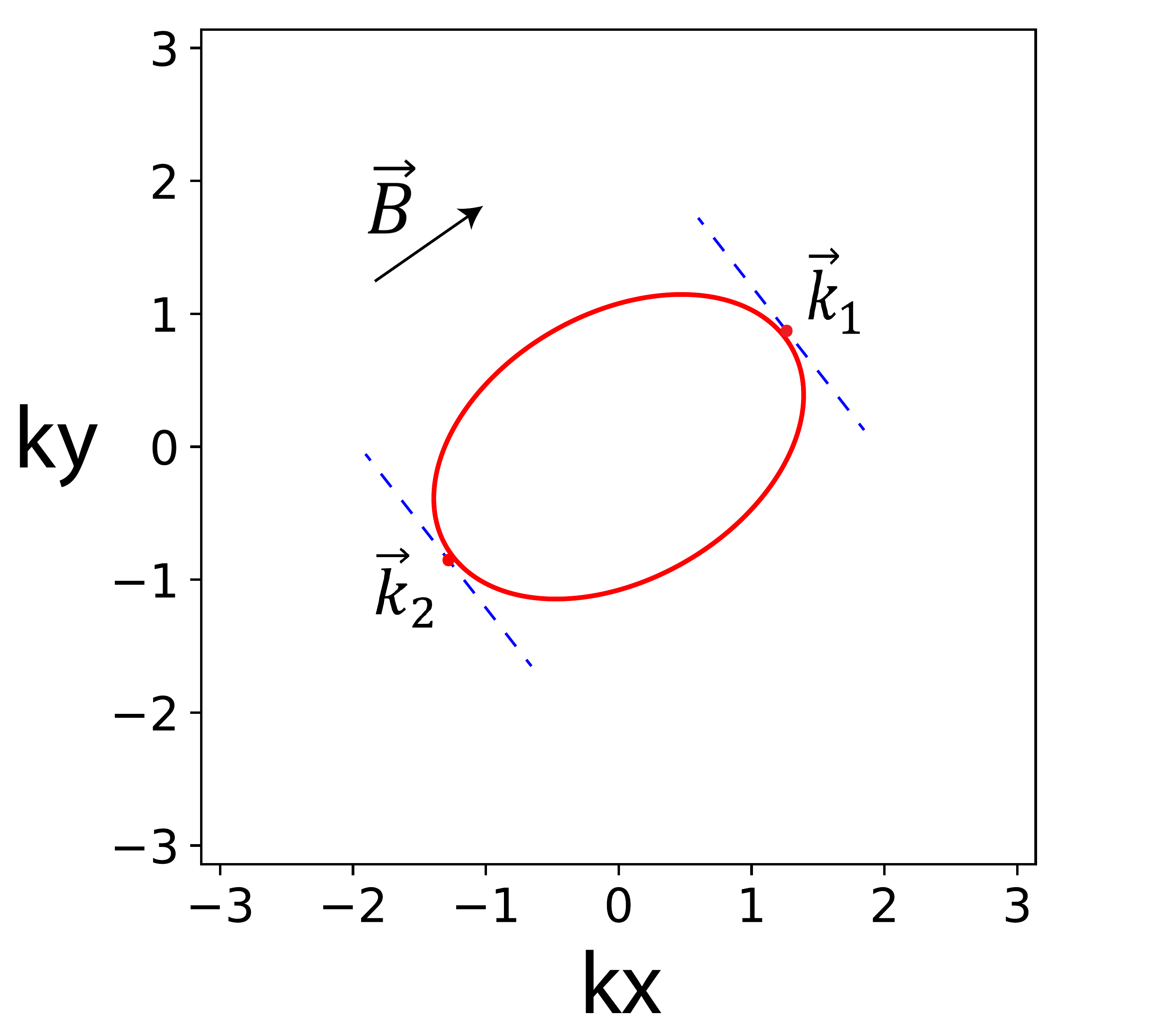}
\caption{The optimal direction of the magnetic field $\bm{B}$ for an anisotropic nodal line in the $k_{x}$-$k_{y}$ plane is to maximize the range $|\bm{k}_{1}-\bm{k}_{2}|$ parallel to $\bm{B}$, which yields the largest $\bar{n}_{D}=|\bm{k}_{1}-\bm{k}_{2}|/\pi$.}
\label{nodes}
\end{figure}

A similar analysis yields the favorable $\bm{q}\equiv \bm{B}$ direction: we note that under the circumstance of the filling of the zeroth Landau level, $\bm{B}$, and thus the energy penalty to the Fermi sea is minimal when $\bar{n}_{D}$ is maximum, which depends on the geometry of the nodal lines and favors the direction with the largest projection. In Fig. \ref{nodes}, we illustrate an example of a simple-loop NL and the direction to maximize the span of the projection $|\bm{k}_{1}-\bm{k}_{2}|$ parallel to $\bm{B}$ and thus $\bar{n}_{D}=|\bm{k}_{1}-\bm{k}_{2}|/\pi$.

\subsection{Our approximations}
Our arguments are valid when the magnetic field is small enough to treat the Dirac nodes of the same $k_{\parallel}$ independently. Otherwise, quantum tunneling kicks in and gaps the Dirac nodes out, which happens if the separation between them $|\Delta{k}_{\perp}|\lesssim|l_{B}^{-1}|$, where $l_{B}=\sqrt{\hbar/eB}$ is the magnetic length \cite{PatrickDFannih, Ramshaw2018}. Those zeroth LLs split beyond the nonzero LLs should not count in Eq. \ref{analysis}. As we can see in Fig. \ref{nodal}a, there are always regions where pairs of Dirac nodes get arbitrarily close and violate the condition $|\Delta{k}_{\perp}|\gg |l_{B}^{-1}|$. Fortunately, the splitting $\propto B$ is generally smaller than the spacing between the zeroth and nonzero LLs $\propto \sqrt{B}$ for small $B$. 

On the other hand, even the Dirac nodes have pairwise-annihilated and developed a mass, as the points $i$ and $j$ in Fig. \ref{nodal}b, the original zeroth LLs may have not yet shifted beyond the first LLs, leading to an underestimation (over-estimation) of $\bar{n}_{3D}$ ($\bm{q}_{opt}$) following Eq. \ref{analysis}. We find the latter effect more dominant in our examples and provide a detailed analysis of these effects in the Appendix A. 

Also, we have assumed that the energy of the Fermi sea depends monotonically on the strength of the magnetic field and is insensitive to its direction; see supporting materials in the Appendix B. 

\begin{figure}
\centering
\includegraphics[width=0.9\linewidth]{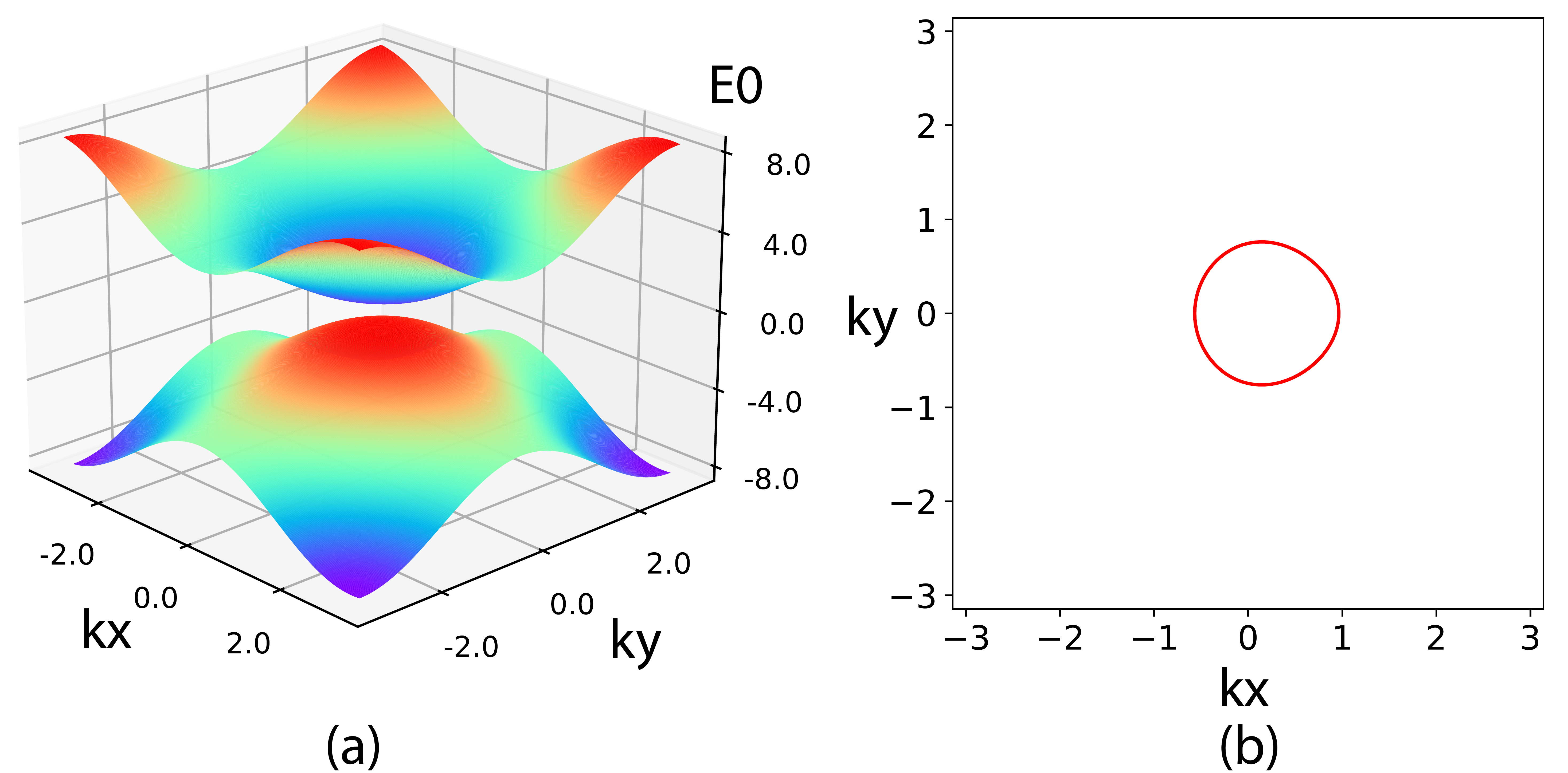}
\caption{(a) The dispersion of $\hat{H}_0$ in Eq. \ref{eq:2Dham} shows shows a $4\epsilon_{0}$ gap between the two bands. $t=1$ ,$t'=0.1$, $\epsilon_{0}=0.5$, $\epsilon_{1}=-2.15$. (b) The Fermi surface of (a) at Fermi energy $E_{F}=1.3$ slightly above half filing shows no sign of FSN.}
\label{fig:disperFermi}
\end{figure}

\section{Numerical results}
\subsection{2D model example}
For demonstration purpose, let's consider a specific effective SDW model $\hat{H}_{2D}=\hat{H}_{0}+\hat{H}_{DW}$ which comes from Eq.\ref{SH2} as follows:
\begin{eqnarray}
\hat{H}_{0}&=&\sum_{\bm{r},\bm{\delta},s,s'}t_{\delta}\hat{c}_{r,s}^{\dagger}\sigma_{s,s'}^{x}\hat{c}_{r+\delta,s'}+\hat{c}_{r,s}^{\dagger}(\boldsymbol{\epsilon}\cdot\boldsymbol{\sigma})_{s,s'}\hat{c}_{r,s'}+h.c. \nonumber \\
\hat{H}_{DW}&=&-\sum_{\bm{r},s,s'}[2V\cos(\bm{q}\cdot\bm{r}+\phi_0)\hat{c}_{r,s}^{\dagger}\sigma_{s,s'}^{z}\hat{c}_{r,s'} \\
&&\label{eq:2Dham}+ 2\lambda \sin(\bm{q}\cdot\bm{r}+\phi_0)\hat{c}_{r,s}^{\dagger}\sigma_{s,s'}^{x}\hat{c}_{r,s'}], \nonumber
\end{eqnarray}
where $\boldsymbol{\sigma}=(\sigma^{x},\sigma^{y},\sigma^{z})$ are the Pauli matrices, $\boldsymbol{\epsilon}=(\epsilon_{1},0,\epsilon_{0})$ are the onsite potentials, and $t_{\delta}=t$ for $\bm{\delta}=\hat{x}, \hat{y}$ and $t_{\delta}=it'$ for $\bm{\delta}=\hat{x}+\hat{y},\hat{x}-\hat{y}$ are the hopping parameters. The DW term $\hat{H}_{DW}$ is similar to the spiral SDW proposed by Overhauser\cite{PhysRev.128.1437}. The dispersion of the translation invariant Hamiltonian $\hat{H}_{0}$:
\begin{equation}
E_k^{0}=\pm 2\sqrt{[\epsilon_{1}+t(\cos{k_x}+\cos{k_y})+2t'\sin{k_x}\cos{k_y}]^{2}+\epsilon_{0}^{2}},
\end{equation}
is shown in Fig. \ref{fig:disperFermi}a, where the Fermi surface slightly above half-filling, as shown in Fig. \ref{fig:disperFermi}b, is rather circular and shows no obvious FSN. However, we will show that to minimize energy, the model prefers a DW, characterized by the term $H_{DW}$, with a preferential wave vector $|\bm{q}_{opt}|$ that shows a continuous range.

\begin{figure}
\centering
\includegraphics[width=0.9\linewidth]{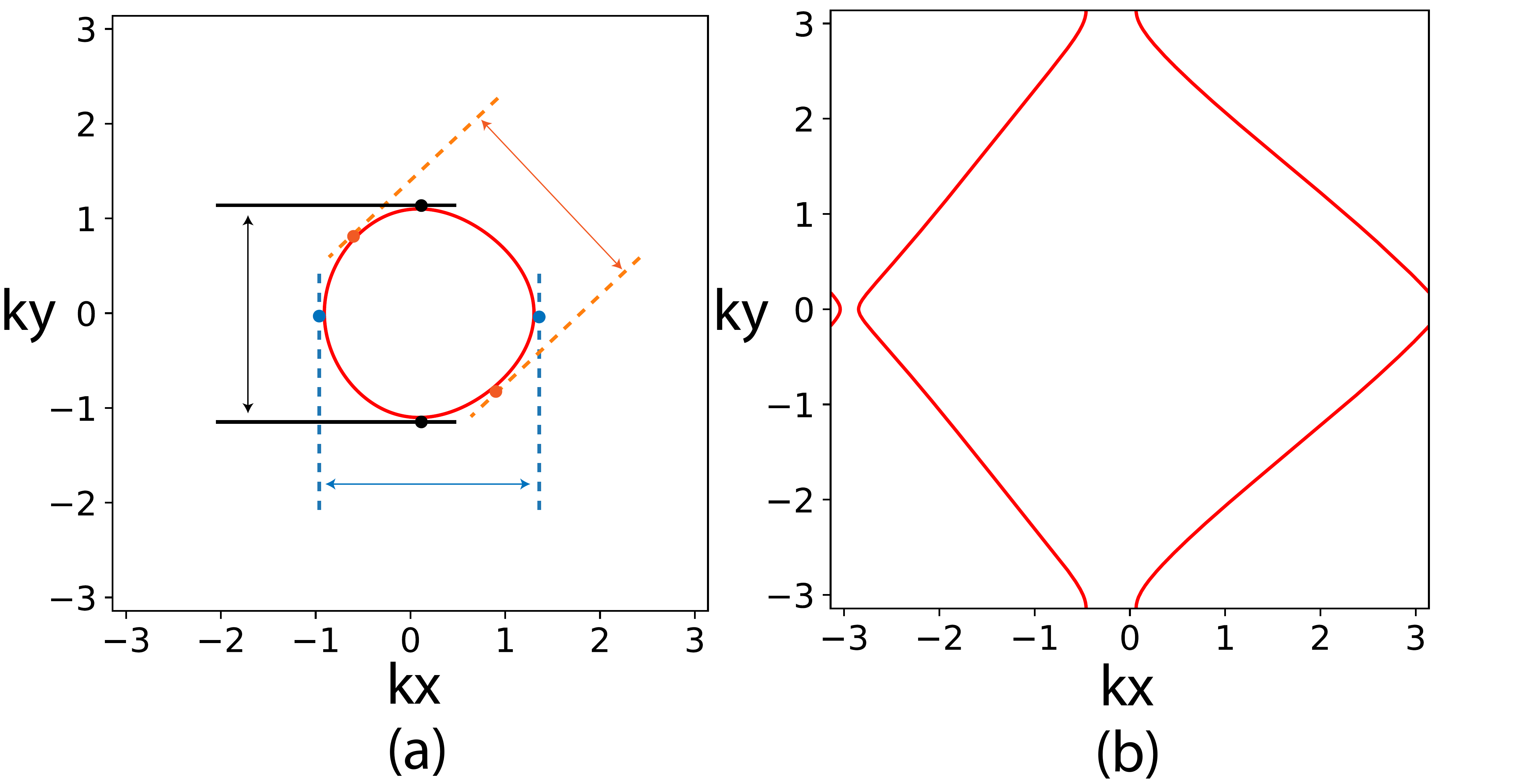}
\caption{The nodal lines of the 3D models in Eq. \ref{eq:disper-3d} in the $k_x - k_y$ planes at $k_{z}=- \frac{\pi}{3}$ for (a) $\lambda=0.8$ and (b) $\lambda=2.5$ close to a Lifshitz transition. $t=1$, $t'=0.1$, $\epsilon_{0}=0.5$, $\epsilon_{1}=-2.15$ as in Fig. \ref{fig:disperFermi}, and $V=1$. The double-sided arrows denote the spans of the nodal lines projected along different directions of the magnetic field $\
bm{B}$.}\label{fig:nodal-line}
\end{figure}

\begin{figure}
\centering
\includegraphics[width=0.9\linewidth]{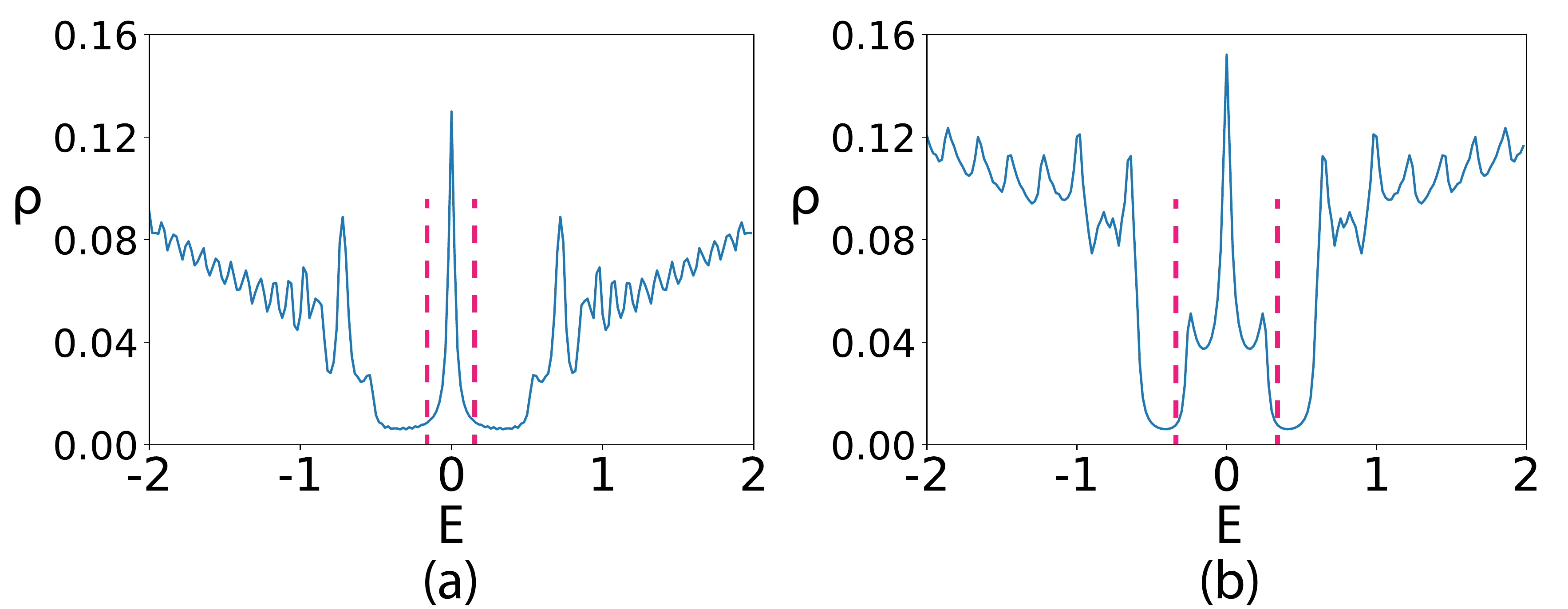}
\caption{The DOS of the 2D system $H_{2D}$ in Eq. \ref{eq:2Dham} with a DW wave vector $\bm{q} = (0.1,0)$ shows a peak around zero energy theoretically attributed to the zeroth LLs of the 3D NL systems in Eq. \ref{eq:disper-3d} in the presence of a magnetic field. The rest of the parameters are the same as in Fig. \ref{fig:nodal-line}. The integrated DOS in (b) is larger given the extent of its NLs. Also, we see signatures of the zeroth LLs' splittings, which remain small compared to the gaps and keep the validity of our argument thanks to the smallness of the magnetic field.}\label{fig:DOS}
\end{figure}

\subsection{3D nodal-line system and LLs}
For an incommensurate $\bm{q}$, the 2D model in Eq. \ref{eq:2Dham} is equivalent to a 3D system with a magnetic field. Without the magnetic field, the 3D system $(\hat{a}_{k,\uparrow}^{\dagger},\hat{a}_{k,\downarrow}^{\dagger})h_{3D}(\bm{k})(\hat{a}_{k,\uparrow},\hat{a}_{k,\downarrow})^{T}$ in the momentum space:
\begin{equation}
    \begin{split}
        &h_{3D}(\bm{k})=[2\epsilon_{1}-2\lambda\sin k_{z}+2t(\cos k_{x}+\cos k_{y})\\
        &+4t'\sin k_{x}\cos k_{y}]\sigma^{x}+[2\epsilon_{0}-2V\cos k_{z}]\sigma^{z},\\
    \end{split}
    \label{eq:disper-3d}
\end{equation}
may possess NLs where the $\sigma^z$ coefficient in Eq. \ref{eq:disper-3d} vanishes on the $k_z= \pm \arccos(\epsilon_0/V)$ planes. We show a couple of examples in Fig. \ref{fig:nodal-line}.

In the presence of the magnetic field, the Dirac nodes along the nodal lines exhibit themselves as Dirac-fermion LLs. While the $n\neq 0$ LLs depend on Fermi-velocity details, the zeroth Landau levels remain (nearly) degenerate at zero energy and are separated from the rest of the Landau bands with large gaps due to the LL spacings. For example, we show in Fig. \ref{fig:DOS} the density of states (DOS) of the models with the NLs in Fig. \ref{fig:nodal-line} in a magnetic field $|\bm{q}|\ll 2\pi$, where the contributions from the zeroth LLs are clearly visible between the red dashed lines.

\subsection{Optimal DW wave vectors}
\begin{figure}
\centering
\includegraphics[width=0.9\linewidth]{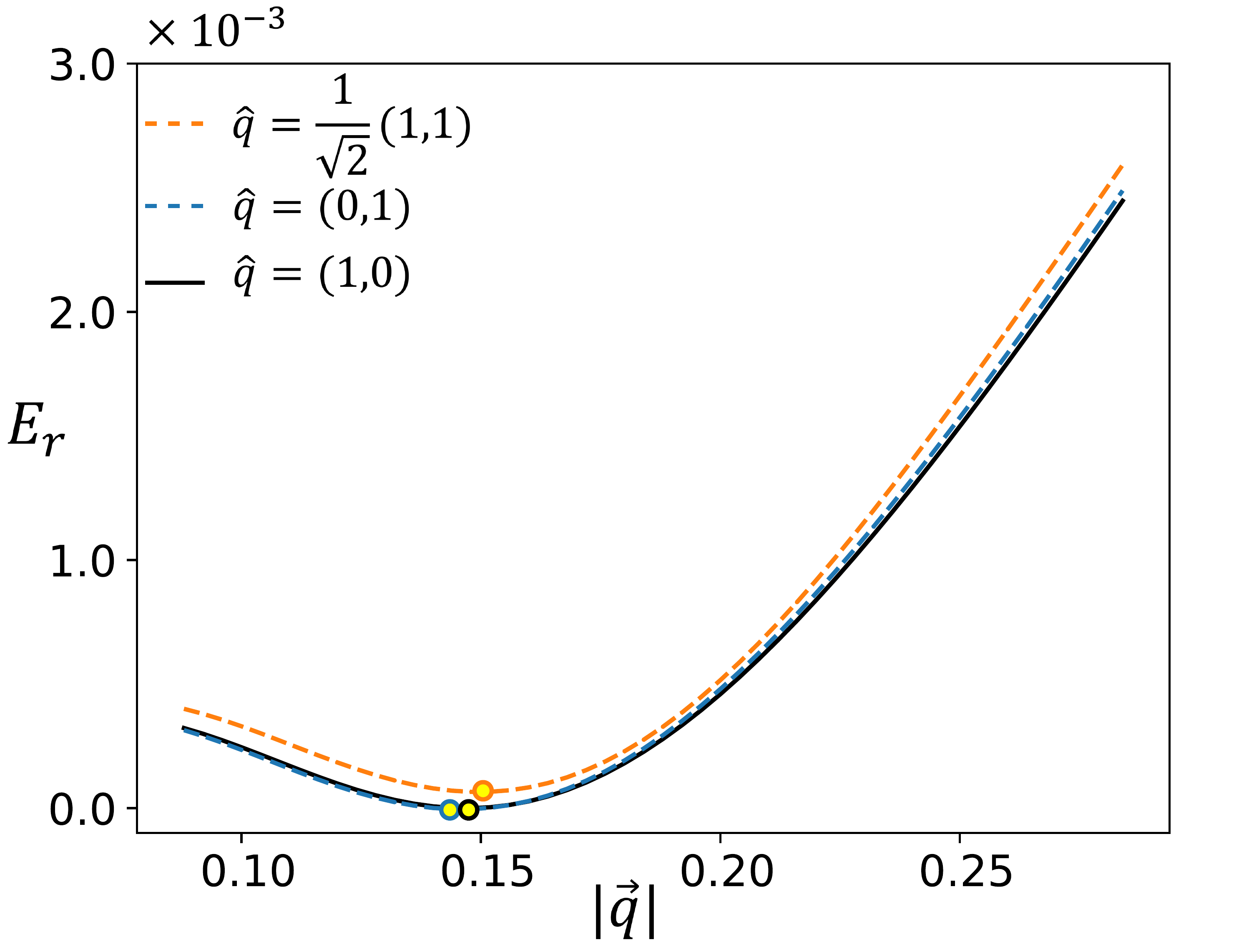}
\caption{The (relative) average energy per electron ($E_r$) $\bar{E}$ for the model in Eq. \ref{eq:2Dham} shows minimums at different optimal $|\bm{q}_{opt}|$ along different directions. The parameters are the same as in Figs. \ref{fig:nodal-line}a and \ref{fig:DOS}a: $t=1$, $t'=0.1$, $\epsilon_{0}=0.5$, $\epsilon_{1}=-2.15$, $V=1$, $\lambda=0.8$, and we set $n_e=1.01$ slightly above half filling. The system has a $y \rightarrow -y$ reflection symmetry thus $\hat q = (\hat x \pm \hat y)/\sqrt{2}$ directions are equivalent.} \label{fig:energyversusq}
\end{figure}

We calculate the energy of $\hat{H}_{2D}$ in Eq. \ref{eq:2Dham} numerically via exact diagonalization on system size $L_{x}=1000$ along $\bm{q}$ and $L_{\perp}=200$ values of $k_\perp$ in the perpendicular direction. We discuss our setup for $\bm{q}$ along directions other than $\hat{x}$ and $\hat{y}$ in the Appendix E. The results on the average energy per electron $\bar{E}$ versus the DW wave vector $\bm{q}$ for the model parameters in Figs. \ref{fig:nodal-line}a and \ref{fig:DOS}a is shown in Fig. \ref{fig:energyversusq}. The optimal wave vector $|\bm{q}_{opt}|\sim 0.15$ is approximately consistent with our theoretical expectation of $|\bm{q}_{theory}|\sim 0.17$ according to Eq. \ref{analysis}. In addition, the DWs in the $\hat{q}=\hat{x},\hat{y}$ directions yield the lowest energy overall with negligible difference, consistent with our analysis and the fact that the NLs projection $|\bm{k}_1-\bm{k}_2|$ and subsequently $\bar{n}_{D}$ are largest along these two directions. In comparison, the average electron energy without the DW $\hat{H}_{DW}$ is $\Delta\bar{E}_0 \sim 0.29$ above these minimums, suggesting that the DW formation is indeed favorable energetically. 

Also, the dependence of $\bar{E}$ with $|\bm{q}|$ is consistent with our expectation: the optimal $|\bm{q}_{opt}|$ allows all the electrons above charge neutrality to be accommodated by the degenerate zeroth LLs; when $0<|\bm{q}|<|\bm{q}_{opt}|$, the filling goes to some of the higher LLs leading to higher energy; when $|\bm{q}|>|\bm{q}_{opt}|$, on the other hand, the electron Fermi sea suffers an energy penalty due to the higher magnetic field. We discuss more detailed results and analysis on energetics in the Appendix B. 
\begin{figure}
\centering
\includegraphics[width=0.9\linewidth]{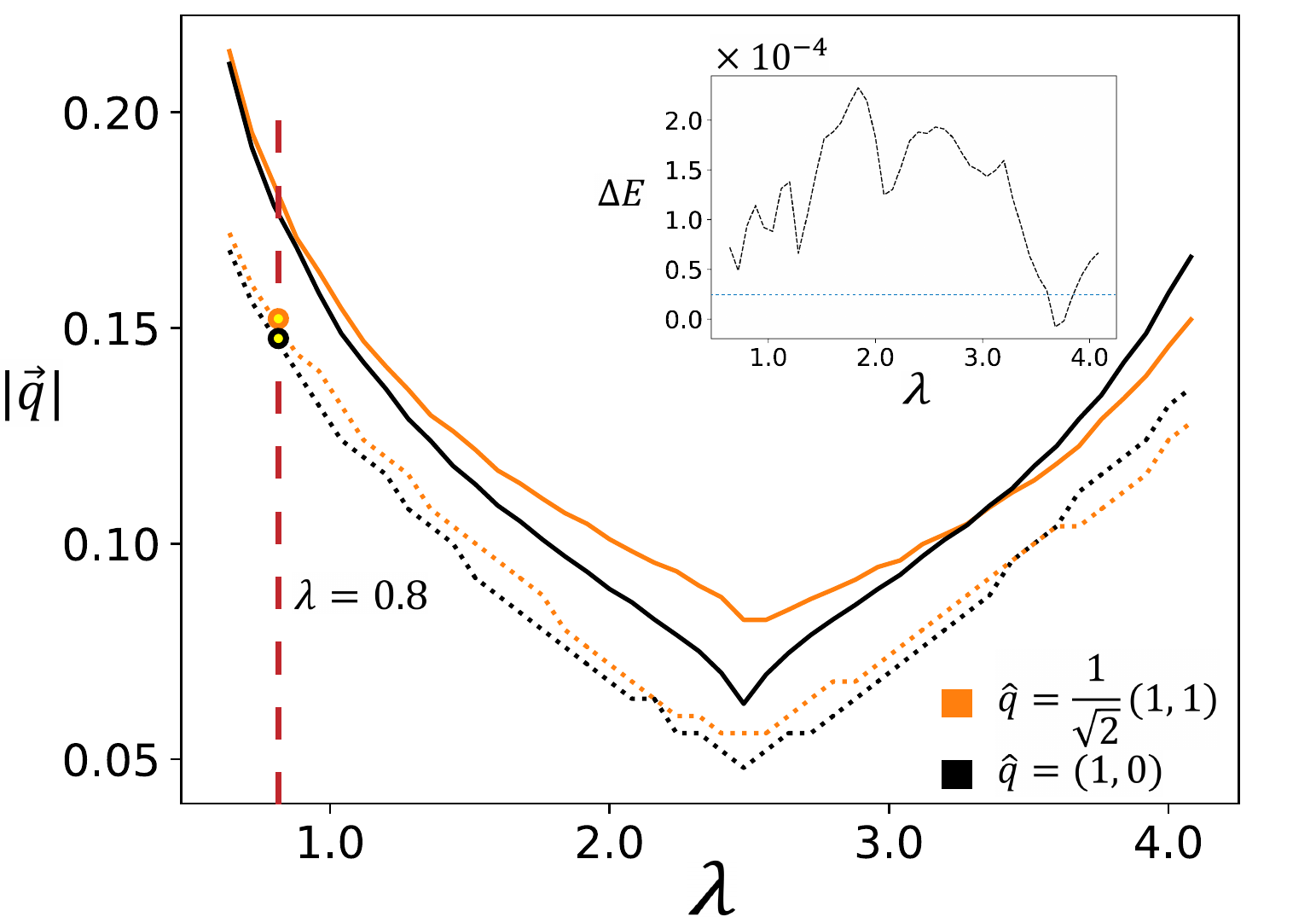}
\caption{The preferential DW wave vector $\bm{q}_{opt}$ with the minimal energy shows good consistency between the numerical results (dashed lines) on the 2D model in Eq. \ref{eq:2Dham} and the theory expectations (solid lines) from Eq. \ref{analysis} and the geometry of the NLs versus the DW strength $\lambda$. The former is obtained by comparing the energy for different $q$ as in Fig. \ref{fig:energyversusq}. Different colors denote results for $\bm{q}$ along the $\hat x$ and the $(\hat x + \hat y)/\sqrt{2}$ directions, and the inset shows the energy difference $\Delta{E}=E_{\frac{1}{\sqrt{2}}(1, 1)}-E_{(1, 0)}$ between the two directions at their respective $\bm{q}_{opt}$. We keep parameters $t=1$, $t'=0.1$, $\epsilon_{0}=0.5$, $\epsilon_{1}=-2.15$, $V=1$, and $n_e=1.01$ as before while varying $\lambda$.}\label{fig:results}
\end{figure}

More generally, the value of $\bm{q}_{opt}$ versus the DW amplitude $\lambda$ as a tuning parameter is summarized in Fig. \ref{fig:results}, and we observe a good consistency between our analysis based on Eq. \ref{analysis} and the numerical benchmark for both $\bm q$ along the $(1,0)$ and the $(1,1)/\sqrt{2}$ directions. Intuitively, when $\lambda$ is small, the nodal line is a closed contour and grows with $\lambda$, leading to an increasing $\bar{n}_{D}$ and decreasing $\bm{q}_{opt}$. On the contrary, as $\lambda$ increases further and beyond the Van Hove singularities, see Fig. \ref{fig:nodal-line}b, the NLs' extent and thus $\bar{n}_{D}$ reverses its trend and decreases until it vanishes, giving rise to a monotonically increasing $|\bm{q}_{opt}|$. The slight differences between numerics and theory are likely due to the neglection of massive Dirac fermion Landau levels (Fig. \ref{nodal}b) that underestimates $\bar{n}_{D}$ (overestimates $\bm{q}_{opt}$), see more related discussions in the Appendix A. In addition, the optimal $|\bm{q}_{opt}|$ is smaller for $\bm{q}$ along the $(1,0)$ direction until large $\lambda \sim 3.5$, suggesting that qualitatively, so should the DW along $\hat{x}$ be the winner with lower energy, which is indeed the case as we compare the two in the inset of Fig. \ref{fig:results}. We also illustrate another example with multiple nodal lines from multiple $k_z$ planes in the Appendix C. Overall, we note that the DWs wave vectors change continuously with the DW parameters, in sharp contrast to the behaviors descending from the FSN or exciton condensation, where the DW wave vector is generally fixed by the Fermi surface geometry.

\section{Conclusions}
In summary, we put forward a novel perspective to understand the DW tendency in 2D systems from the energetics of a 3D NL system. Interestingly, in scenarios where DW sees a clear FSN origin, our perspective also offers a consistent explanation; see the Appendix D. On the other hand, our setup goes beyond the Peierls transition as it does not require any apparent FSN to begin with. Correspondingly, the optimal SDW wave vector depends on the geometry of the 3D NLs and may vary continuously with respect to the DW parameters without changes to the Fermi surface and dispersion. Also, our numerical results on benchmark models fit with our analysis consistently. Such continuous variations of the DW wave vectors are present in materials such as various cuprates \cite{PhysRevX.9.021021, Hoffman,sci5584, RN1}, $2H$-Ta${\mathrm{Se}}_{2}$\cite{MC1, MC2} etc. While we do not intend to relate our analysis to these materials directly, our perspective kindles the theoretical possibility of a variable DW wave vector from more generic origins.

On the other hand, our study points out that our current understanding of the DW origin is still primitive and a universal understanding is not yet available. Our current study has focused on models with rather specific constructions. While such a setup facilitates the theoretical analysis and its controllability, it also limits the generalization of the mechanism and its application in practice. It will be interesting to probe the generalization of the current mechanism beyond its model limits and its connection with FSN and other DW mechanisms in interpolating models for further physical intuition. 

Finally, we note that our theory offers a bridge between density waves and topological nodal-line semi-metal. It is also interesting to explore the topological property of the density wave system via the dimension-extension perspective \cite{PhysRevB.91.014108}.

\section{Acknowledgement}
We thank Xin-chi Zhou and Di-zhao Zhu for insightful discussions. The authors are supported by the National Science Foundation of China(No.12174008) and the start-up grant at Peking University. The calculations of this work is supported by HPC facilities at Peking University.

\bibliography{ref.bib}
\section{Appendix A: The independent-Dirac-node approximation}

In the main text, we regard the Dirac nodes as independent for our derivations. However, quantum tunneling exists between the Dirac nodes, especially in the presence of a larger magnetic field. Here, we perform a relatively quantitative study on the potential impact of the Dirac-node pair separation and the magnetic field strength. 

We consider the following two-bands model to account for the low-energy scenarios at different $k_\parallel$ cross-sections in Fig. 1 in the main text:
\begin{equation}
\hat{H}_{D} = (m-k_{x}^2)\sigma^{x}+k_{y}\sigma^{y},
\label{two_node}
\tag{A1}
\end{equation}
where $m$ is a parameter controlling whether the model has two massless Dirac-fermions at $( \pm \sqrt{m},0)$ ($m>0$) or massive Dirac-fermions with a mass $\sim |m|$ ($m<0$).

\subsection{1. Dirac-node pair separation}

First, we consider the $m>0$ case similar to the specific $k_\parallel$ cross-sections intersecting the nodal line in Fig. 1 in the main text. Different $m$ corresponds to different Dirac-node separations, which affect our theory's accuracy.

\begin{figure}
\centering
\includegraphics[width = 0.9\linewidth]{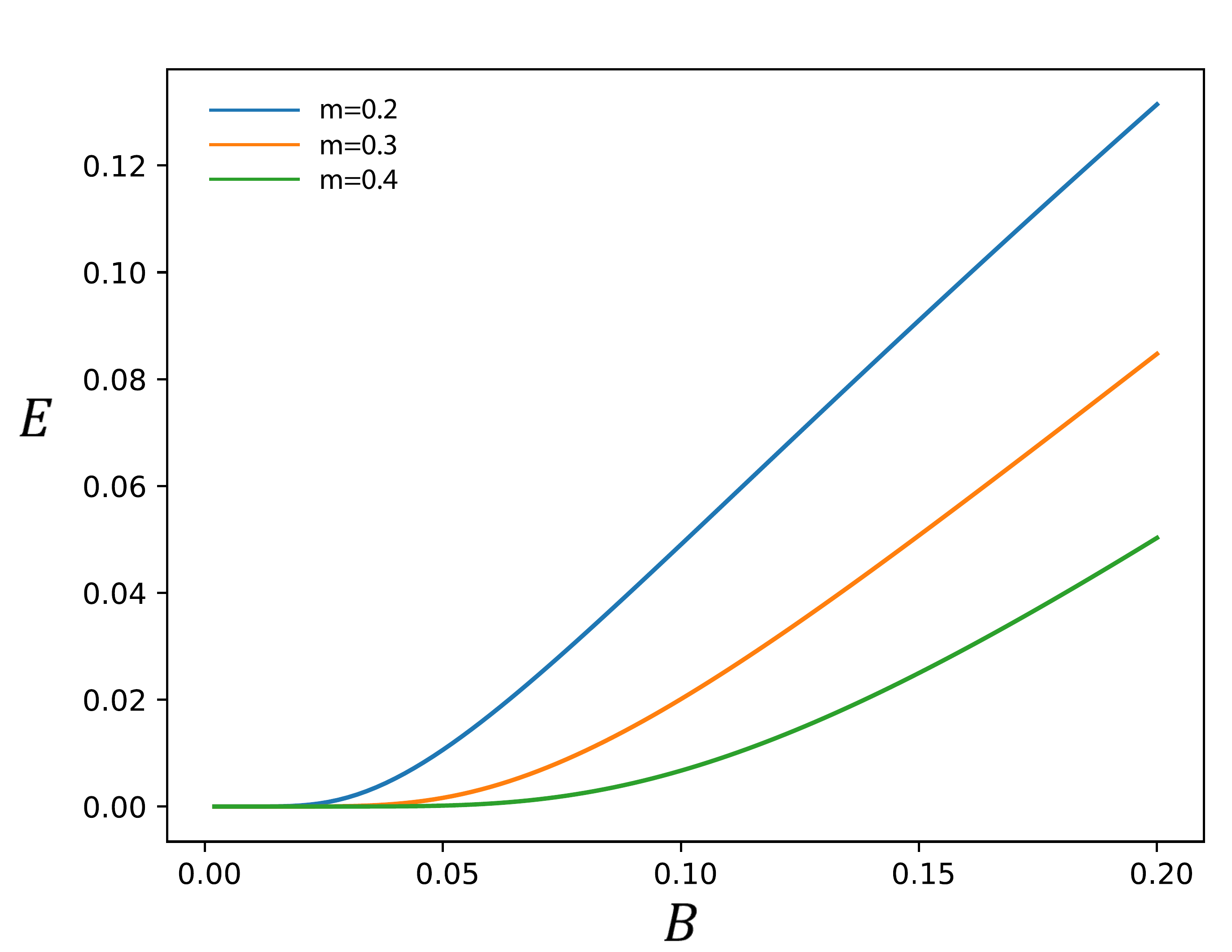}
\caption{The zeroth LL drifts away from zero energy as the magnetic field perpendicular to the Dirac-node separation increases, especially when the inverse magnetic length $l_{B}^{-1}\sim \Delta k_{\perp}=2\sqrt{m}$. For instance $\Delta k_{\perp}\approx 1.26$ for $m=0.4$, and a visible energy drift appears at $B\sim 0.1$, $l_{B}^{-1} \approx 0.25\Delta k_{\perp}$.}
\label{Diracgap}
\end{figure}

We can focus on one of the Dirac node $\chi = \pm 1$ by $k_{x}'=k_{x}-\chi\sqrt{m}, k_{y}'=k_{y}$, and the corresponding model becomes:
\begin{equation}
\hat{H}_{\chi}=v_{x}k_{x}'\sigma^{x}+v_{y}k_{y}'\sigma^{y}-k_{x}'^{2}\sigma^{x},
\tag{A2}
\end{equation}
where the Fermi velocities are $v_{x}=-2\chi\sqrt{m}$ and $v_y=1$. When we apply a magnetic field $\bm{B}=(0,0,B)$ with the gauge $\bm{A}=(0,Bx,0)$, we can re-express $k_{x}'$ and $k_{y}'$ in terms of ladder operators $\hat{a}$ and $\hat{a}^{\dagger}$:
\begin{equation}
    \begin{split}
        &k_{x}'=\sqrt{\frac{v_{y}}{|v_{x}|}}\frac{1}{\sqrt{2}l_{B}}(\hat{a}+\hat{a}^{\dagger}),\\
        &k_{y}'+\frac{eBx}{\hbar}=\sqrt{\frac{|v_{x}|}{v_{y}}}\frac{1}{\sqrt{2}l_{B}}i(\hat{a}-\hat{a}^{\dagger}),\\
    \end{split}
    \tag{A3}
\end{equation}
where $l_B=\sqrt{\frac{\hbar}{eB}}$ is the magnetic length. For instance, the Hamiltonian for $\chi=+1$ becomes:
\begin{equation}
    \hat{H}_{+1}=-\frac{\sqrt{2|v|_{x}v_y}}{l_{B}}(\hat{a}\sigma^{-}+\hat{a}^{\dagger}\sigma^{+})-\frac{v_{y}|v_{x}|}{4ml_{B}^{2}}(\hat{a}+\hat{a}^{\dagger})^{2}\sigma^{x}.
    \tag{A4}
\end{equation}
The first term on the right-hand side corresponds to the Dirac-fermion Landau levels (LLs) of an independent Dirac node and possesses a zeroth LL at precisely zero energy. The second term captures the tunneling from/to the other Dirac node ($\chi=-1$) and introduces hybridization between the LLs. When the separation surpasses the inverse magnetic length $l_{B}^{-1}\ll\Delta k_{\perp}=2\sqrt{m}$, the influence of the second term is negligible. However, when $l_{B}^{-1}\sim\Delta k_{\perp}$,  we can no longer treat the two Dirac nodes independently, and the second term becomes non-perturbative; their zeroth Landau levels interact and split \cite{PatrickDFannih, Ramshaw2018}. Fig. \ref{Diracgap} illustrates the zeroth LL's drift from zero energy as the magnetic field increases. A larger separation ($m$) protects the independent-Dirac-node approximation until a larger magnetic field $B$. 

For a closed nodal line, e.g., the one in Fig. 1 in the main text, there always exist pairs of Dirac nodes with separation $\Delta k_{\perp}\sim l_{B}^{-1}$ on cross-sections close to the turning points regardless of the magnetic field. These Dirac nodes gap out in pairs leading to an overestimate of the zeroth LL density and underestimate of the density wave vector. Still, the less the magnetic field, the fewer corners we cut (metaphorically and literally), and the more robust the approximations. 

\subsection{2. Massive Dirac fermions}

In the main text, we focus solely on the massless Dirac fermions in our formulations. However, when the mass is small, the massive Dirac fermion may behave like a massless one and contribute to the zero-energy density peak, thus the optimal wave vector. 

In this subsection, we focus on the influence of massive Dirac fermions, and consider the model in Eq. \ref{two_node} with $m<0$. In the presence of a magnetic field $\bm{B}=(0,0,B)$ with gauge $\bm{A}=(0,Bx,0)$, we can re-express the momentum in terms of ladder operators $\hat{a}$ and $\hat{a}^{\dagger}$ as:
\begin{equation}
    \begin{split}
        &k_{x} = \frac{1}{\sqrt{2}l_{B}}(\hat{a}+\hat{a}^{\dagger}),\\
        &k_{y}+\frac{eBx}{\hbar} = i\frac{1}{\sqrt{2}l_{B}}(\hat{a}-\hat{a}^{\dagger}),\\
    \end{split}
    \tag{A5}
\end{equation}
and the Hamiltonian as: 
\begin{equation}
\hat{H}_{D'} = \{m-[\frac{1}{\sqrt{2}l_{B}}(\hat{a}+\hat{a}^{\dagger})]^2\}\sigma^{x}+i\frac{1}{\sqrt{2}l_{B}}(\hat{a}-\hat{a}^{\dagger})\sigma^{y}.
\label{h_split}
\tag{A6}
\end{equation}

\begin{figure}
\centering
\includegraphics[width = 0.9\linewidth]{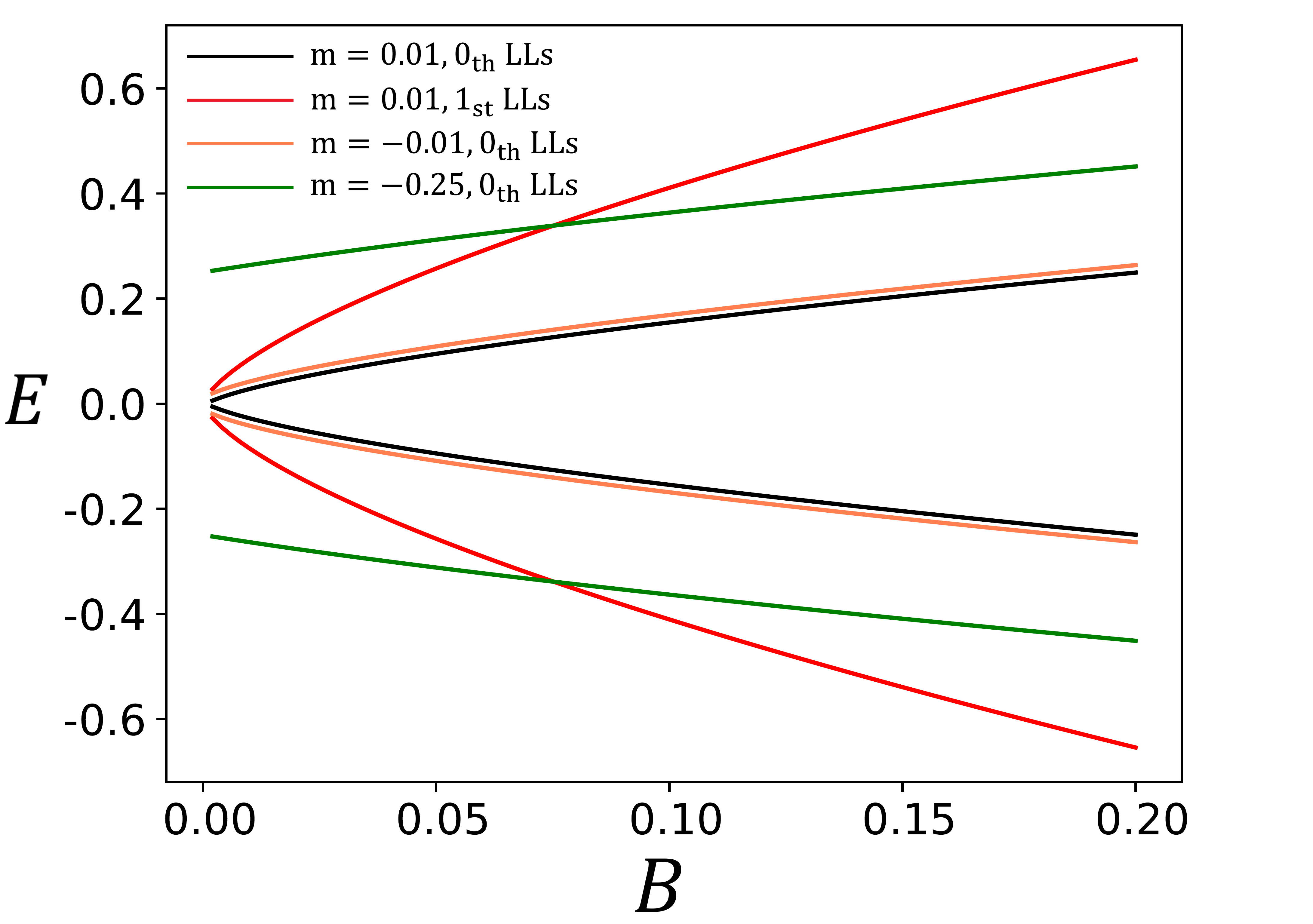}
\caption{Selected LLs of of Eq. \ref{h_split} versus the magnetic field $B$ for massless ($m= 0.01$) and massive ($m=-0.01$, $m=-0.25$) Dirac fermions dictate the possible contributions to the DOS around zero energy.}
\label{gap}
\end{figure}

We illustrate the low-lying LLs with selected model parameters in Fig. \ref{gap}. The results indicate that even when the Dirac nodes have just annihilated in pairs ($m<0$), the resulting LLs may still be relatively close to zero energy depending on the mass $m$. For instance, the zeroth LLs for $m=-0.01$ in Fig. \ref{gap} differs little from that for $m=0.01$. By counting only the nodal lines, we may underestimate the contribution from such (slightly) massive Dirac-fermion LLs to zero-energy density peak, thus the optimal wave vector.

Still, the nodal lines offer a generally good starting point, as the zeroth LLs for most massive fermion LLs, e.g., $m=-0.25$, and the nonzero LLs, are too far from zero energy to contribute; see Fig. \ref{gap}.

\section{Appendix B: Energy cost of magnetic field on the Fermi sea}

\begin{figure}
\centering
\includegraphics[width = 0.9\linewidth]{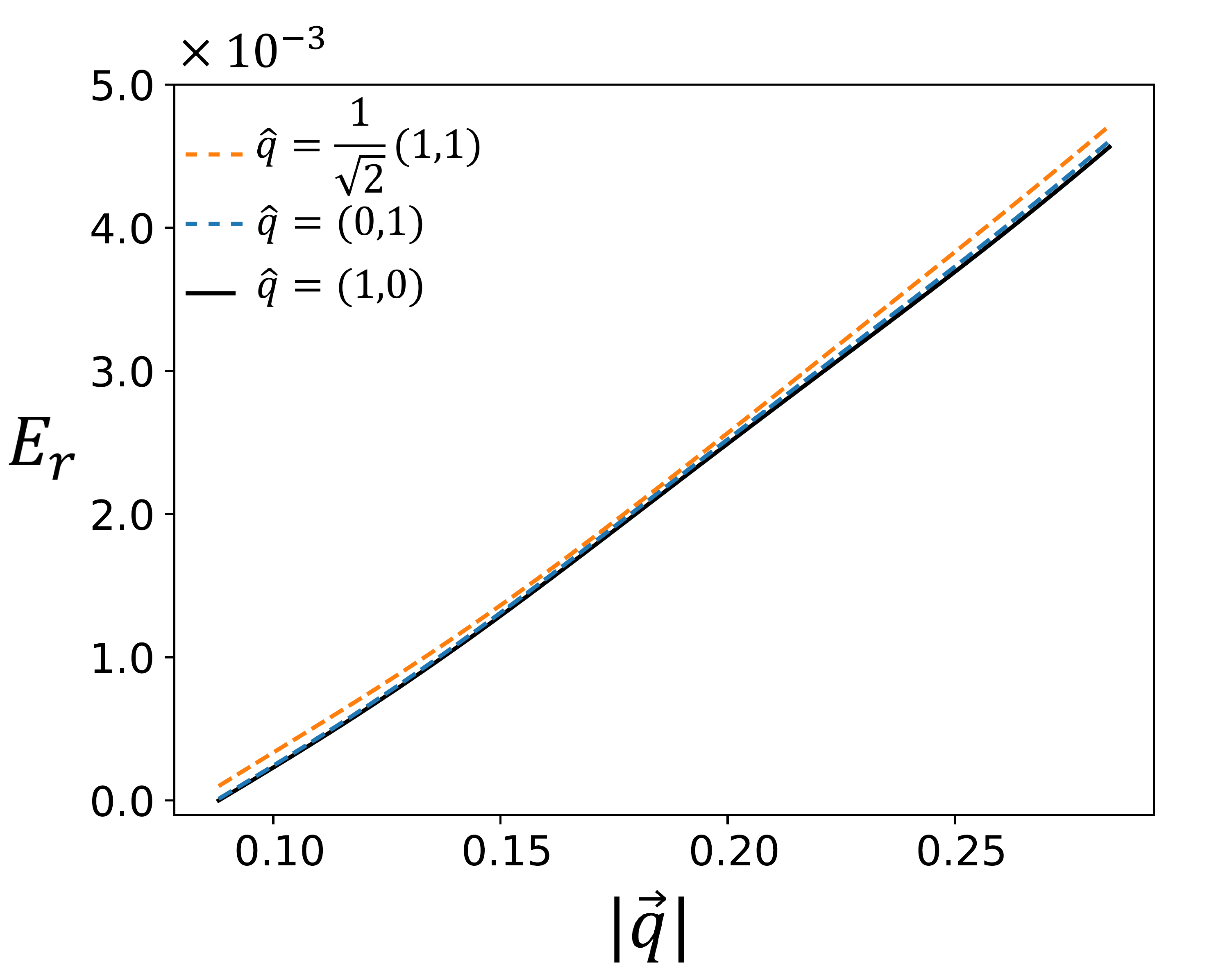}
\caption{The (relative) average energy ($E_r$) $E_0$ of the electrons in the Fermi sea at half-filling $n_e=1$ depends monotonically on the largeness of $\bm{q}$ yet little on the direction of $\bm{q}$. We have applied the same parameter setting as in Figs. 4a and 5a in the main text.}
\label{background}
\end{figure}

We calculate the energy of the Fermi sea, i.e., the average energy of the electrons at half-filling $n_e=1$ for the model in Figs. 4a and 5a in the main text in a magnetic field $B$, and summarize the results in Fig. \ref{background}. The energy increases monotonically with respect to the amplitude of the magnetic field (DW wave vector) and is relatively insensitive to the direction of the magnetic field (DW wave vector). Therefore, we cannot arbitrarily increase $|\bm{q}_{opt}|$ after reaching the complete filling of zeroth LLs due to the associated energy cost. 

From a phenomenological perspective, we may attribute the energy cost of the magnetic field on the Fermi sea to the orbital diamagnetic effect of the electrons. Otherwise, i.e., a small magnetization reduces the energy, the Fermi sea is unstable toward spontaneous magnetization, contradictory with the ground state precondition. Such an argument is valid within the limit of a small magnetic field
where we mainly focus.

\section{Appendix C: Example with multiple nodal lines}

For simplicity, we focus our discussions in the main text on scenarios with a nodal line in a single $k_z$-plane. It is straightforward to generalize our conclusions to multiple nodal lines and over multiple $k_z$-plane, where all contributions add to the zeroth LLs' degeneracy and the eventual optimal wave vector. 

\begin{figure}
\centering
\includegraphics[width = 0.9\linewidth]{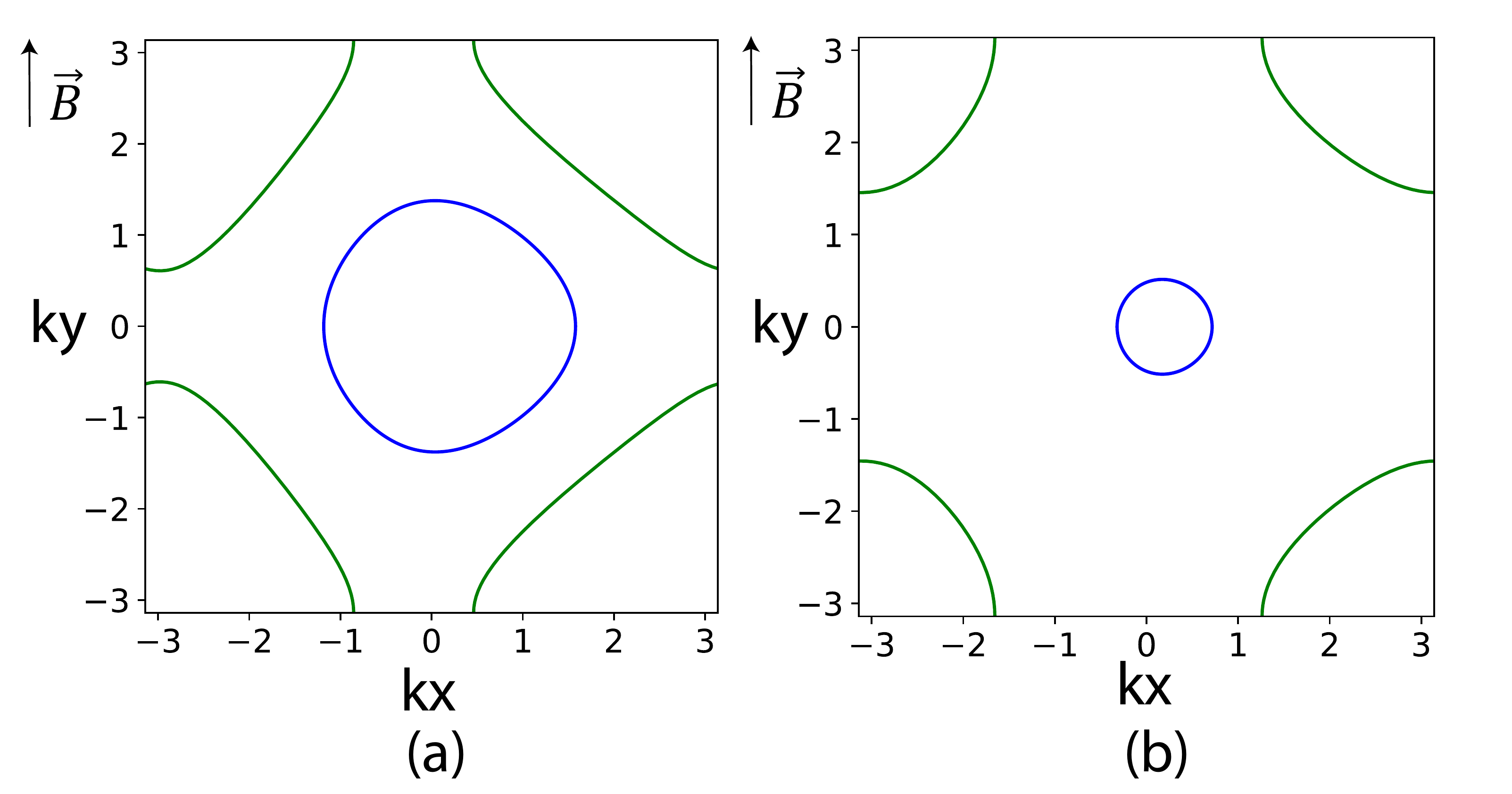}
\caption{The model in Eq. \ref{2-nodal line} possesses two nodal lines in the $k_{z}=\frac{\pi}{3}$ (blue) and $k_{z}=-\frac{\pi}{3}$ (green) planes, respectively, for parameters $t=1$, $t'=0.1$, $\epsilon_{0}=0.5$, $\epsilon_{1}=-0.5$, $V=1$, (a) $\lambda=0.8$, and (b) $\lambda=1.6$. As $\lambda$ grows, the inner nodal line (blue) shrinks while the outer nodal line (green) expands to the corners till it vanishes.}
\label{2nodalline}
\end{figure}

Here, we illustrate such an example. Let's consider the model in Eq. 9 in the main text:
\begin{equation}
    \begin{split}
        &h_{3D}(\bm{k})=[2\epsilon_{1}-2\lambda\sin k_{z}+2t(\cos k_{x}+\cos k_{y})\\
        &+4t'\sin k_{x}\cos k_{y}]\sigma^{x}+[2\epsilon_{0}-2V\cos k_{z}]\sigma^{z}.\\
    \end{split}
    \label{2-nodal line}
    \tag{C1}
\end{equation}
The model may possess two nodal lines as illustrated in Fig. \ref{2nodalline}. 

\begin{figure}
\centering
\includegraphics[width = 0.9\linewidth]{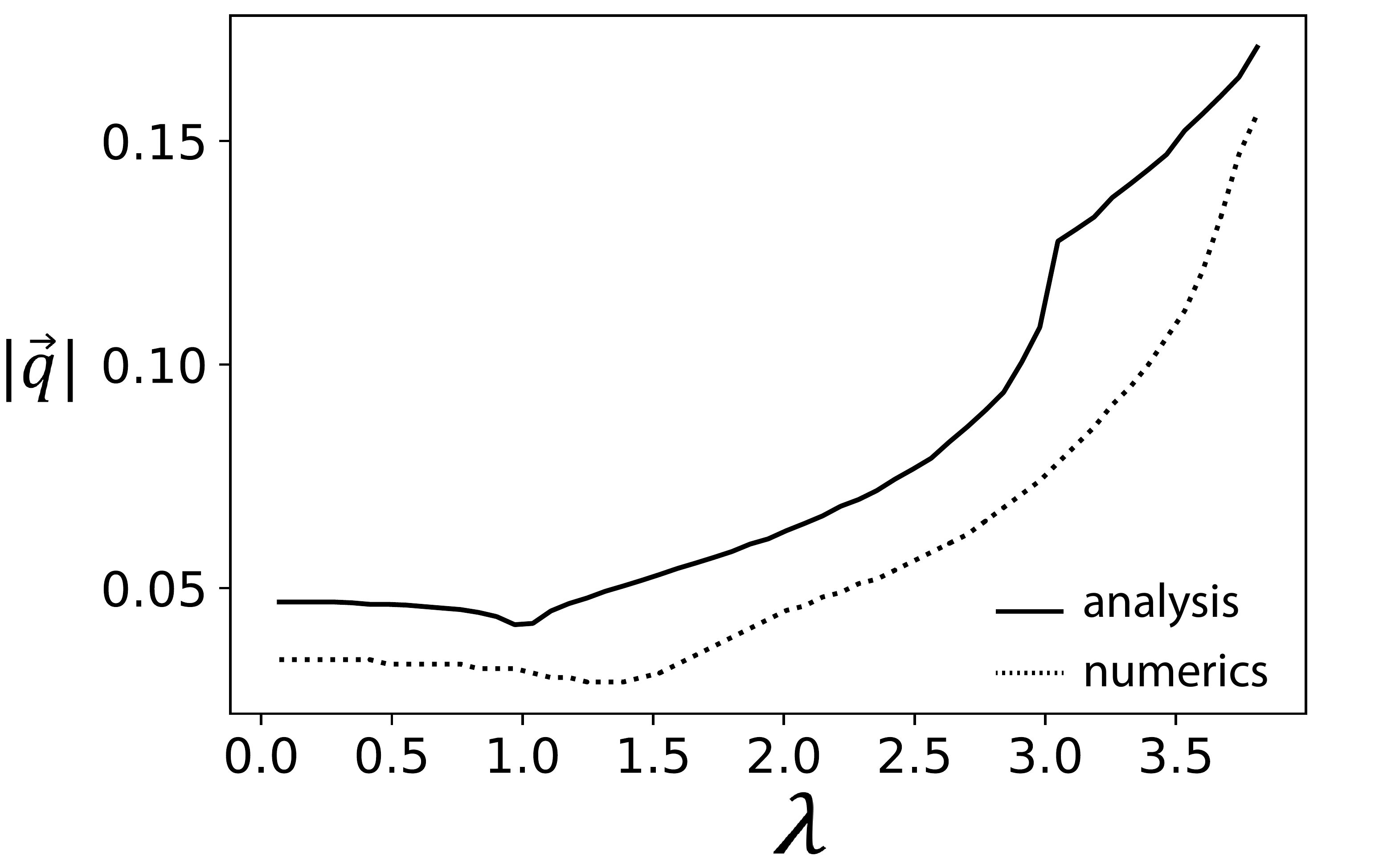}
\caption{The theoretical optimal wave vector $\bm{q}_{opt}$ and the numerical results compare relatively well and exhibit a consistent tendency over a wide range of $\lambda$ parameter range. We carry out numerical calculations on 2D system of size $L_{x}=1000$ and $L_{y}=200$.}
\label{comparereferee}
\end{figure} 

According to our theoretical analysis, the optimal wave vector should follow Eq. 6 in the main text: 
\begin{equation}
 n_e-1 =\frac{\bar{n}_{D}}{2}\frac{|\bm{q}_{opt}|}{2\pi}, \label{eq:expect}
 \tag{C2}
\end{equation}
whose $\bar{n}_{D}$ incorporates both nodal lines from the two different $k_{z}$ planes. The nodal lines possess the largest overall projection along the $\hat{y}$ direction, predicting an optimal magnetic field (wave vector) along the $\hat{B}=(0, 1)$ ($\hat{q}=(1, 0)$) direction. We compare the expected optimal wave vector $\bm{q}_{opt}$ from Eq. \ref{eq:expect} and the numerical results for various values of $\lambda$ in Fig. \ref{comparereferee}, which shows overall consistency.

\section{Appendix D: Consistency with the Fermi surface nesting}

In the main text, we show that our perspective tackles density waves and their optimal wave vectors beyond the Fermi surface nesting (FSN). An interesting question is whether our perspective works for scenarios where FSN is still applicable. In this subsection, we show an example where our theory is consistent with FSN. For instance, let's consider the quasi-1D limit of the model in Eq. 7 in the main text:
\begin{equation}
    \begin{split}
    \hat{H}_{0}&=\sum_{\bm{r},\bm{\delta}}t_{\delta}\hat{c}_{r,s}^{\dagger}\sigma_{s,s'}^{x}\hat{c}_{r+\delta,s'}+\hat{c}_{r,s}^{\dagger}(\boldsymbol{\epsilon}\cdot\boldsymbol{\sigma})_{s,s'}\hat{c}_{r,s'}+\mbox{h.c.},\\
    \hat{H}_{DW}&=-\sum_{\bm{r},s,s'}[2V\cos(\bm{q}\cdot\bm{r}+\phi_0)\hat{c}_{r,s}^{\dagger}\sigma_{s,s'}^{z}\hat{c}_{r,s'}\\
    &+ 2\lambda \sin(\bm{q}\cdot\bm{r}+\phi_0)\hat{c}_{r,s}^{\dagger}\sigma_{s,s'}^{x}\hat{c}_{r,s'}],\\
    \end{split}
    \tag{D1}
\end{equation}
where $\boldsymbol{\sigma}=(\sigma^{x}, \sigma^{y}, \sigma^{z})$ are the Pauli matrices, $\boldsymbol{\epsilon}=(\epsilon_1, 0, \epsilon_0)$ are the onsite potentials, $t_{\delta}=t, 0$ for $\bm{\delta}=\hat{x}, \hat{y}$, and $t_{\delta}=it'$ for $\bm{\delta}=\hat{x}+\hat{y}, \hat{x}-\hat{y}$.

\begin{figure}
\centering
\includegraphics[width = 0.9\linewidth]{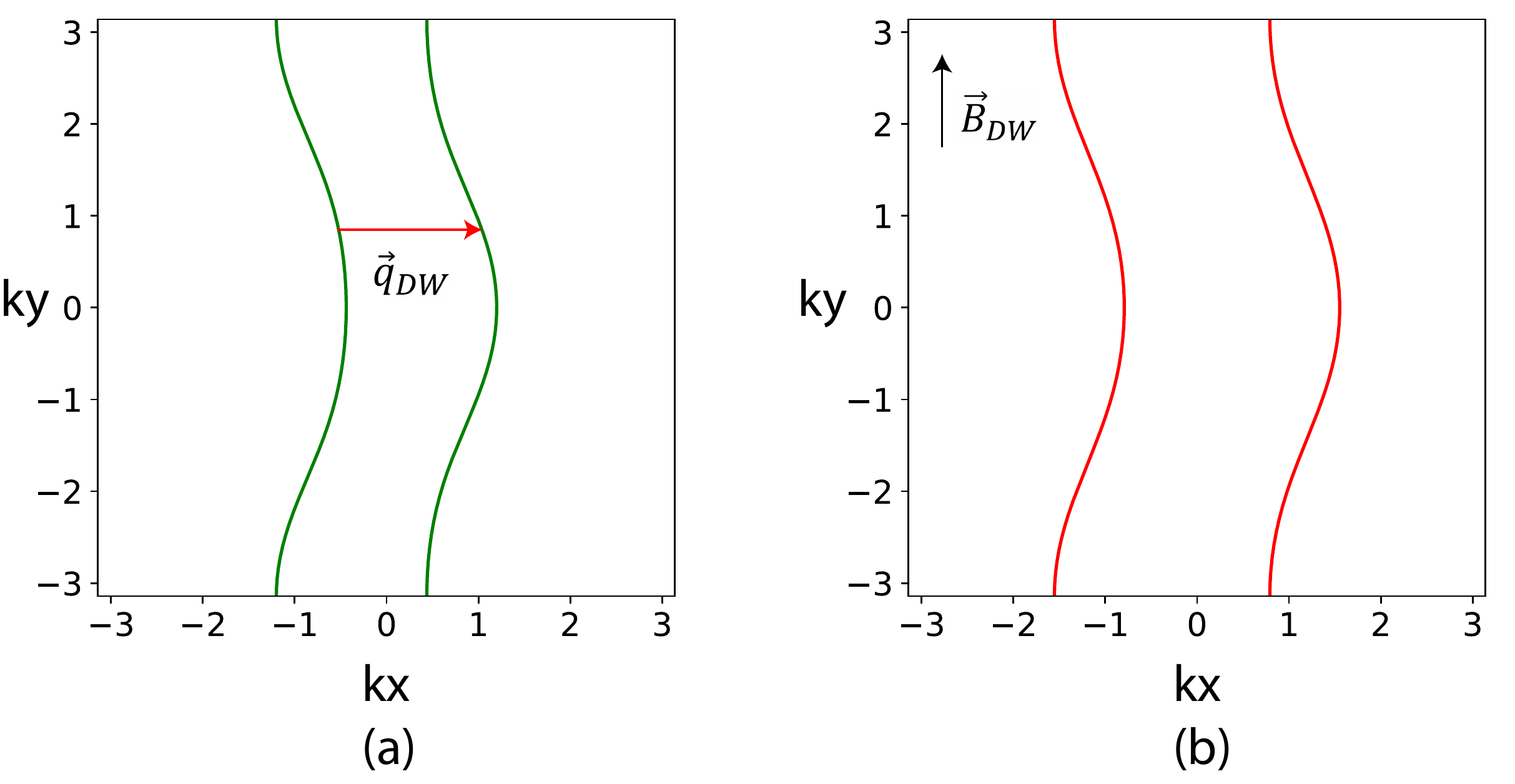}
\caption{(a) The Fermi surface of $h_{0}(\bm{k})$, $\hat{H}_{0}$ in the $\bm{k}$-space, shows a nearly perfect nesting wave vector $\bm{q}_{DW}$. (b) The nodal lines of $h_{3D}(\bm{k})$ prefers the optimal magnetic field $\bm{B}_{DW}$ along the $\hat y$ direction. Here, $t=1$, $t'=0.2$, $\epsilon_{1}=-2.15$, $\epsilon_{0}$, $V=1$, $\lambda=2$, $n_{e}=1.01$}\label{1Dlimit}.
\end{figure}

The Fermi surface of $\hat{H}_{0}$ in the $\bm{k}$-space:
\begin{equation}
    h_{0}(\bm{k})=[2\epsilon_1+2t\cos(k_x)+4t'\cos(k_x)\sin(k_y)]\sigma^{x}+2\epsilon_{0}\sigma^{z}, 
    \tag{D2}
\end{equation}
exhibits nearly perfect FSN between the two pieces as illustrated in Fig. \ref{1Dlimit}a, indicating a density-wave tendency with a preferential wave vector $\bm{q}_{DW}$:
\begin{equation}
    n_{e}\approx 1+\frac{|\bm{q}_{DW}|}{2\pi}.
    \tag{D3}
\end{equation}

On the other hand, as we map the entire 2D system $\hat{H}_{0} + \hat{H}_{DW}$ to 3D, the 3D model takes the form in the absence of a magnetic field:
\begin{equation}
    \begin{split}
         h_{3D}(\bm{k})&=[2\epsilon_1-2\lambda\sin(k_z)+2t\cos(k_x)+\\
         &4t'\cos(k_x)\sin(k_y)]\sigma^{x}+[2\epsilon_{0}-2V\cos(k_z)]\sigma^{z},\\
    \end{split}
    \tag{D4}
\end{equation}
whose nodal lines are shown in Fig. \ref{1Dlimit}b. The optimal magnetic field $\bm{B}_{DW}$ favors the direction with the largest nodal-line projection, which is the $\hat{y}$ direction in this model. Further, we use Eq. \ref{eq:expect} (Eq. 6 in the main text) to derive the optimal density wave vector:
\begin{equation}
    \begin{split}
        (n_{e}-1)L_{\parallel}S_{\perp}&=\frac{\int n(k_{\parallel})dk_{\parallel}}{2\pi/L_{\parallel}}\frac{|\bm{q}_{opt}|}{2\pi}\frac{S_{\perp}}{2}=\frac{4\pi}{2\pi/L_{\parallel}}\frac{|\bm{q}_{opt}|}{2\pi}\frac{S_{\perp}}{2}\\
        &\Rightarrow|\bm{q}_{opt}|=2\pi(n_{e}-1).\\
    \end{split}
    \tag{D5}
\end{equation}

It is straightforward to see the consistency between $\bm{q}_{DW}$ from FSN and $\bm{q}_{opt}$ from our novel theoretical perspective:
\begin{equation}
    |\bm{q}_{opt}|\approx|\bm{q}_{DW}|,
    \tag{D6}
\end{equation}
and both along the $\hat x$ direction. Our theory is capable of incorporating as well as going beyond the conventional FSN\cite{DzYz}.

\section{Appendix E: Numerical method for systems with a tilted magnetic field}

Generally, density waves may arise in any direction, and we need to consider the case where $\bm q=(q_{x},q_{y})$ is tilted from the $\hat x$ or $\hat y$ directions for the most energetically favorable condition. Here is the method we used to compute the dispersion of the model in Eq. 7 of the main text with $\bm{q}$ in any direction.

While our theoretical analysis based upon 3D NLs is a low-energy effective theory that can apply to $\bm {q}$ along any direction, the numerical calculations require that $\bm{q}=(q_{x},q_{y})$ points to a commensurate direction, $\frac{q_{x}}{q_{y}}=\frac{m}{n},m,n\in \mathbb{Z}$, $gcd(m,n)=1$. Without loss of generality, we limit ourselves to: 
\begin{equation}
\bm{q}=|q_{0}|(\frac{1}{\sqrt{1+p^2}},\frac{p}{\sqrt{1+p^2}}),
\tag{E1}
\end{equation}
for simplicity, so that there will not be additional sublattices. Here, $|q_0|$ is the norm of the wave vector, and $p\in \mathbb{Z}$ determines the direction of $\bm{q}$.

\begin{figure}
\centering
\includegraphics[width=0.9 \linewidth]{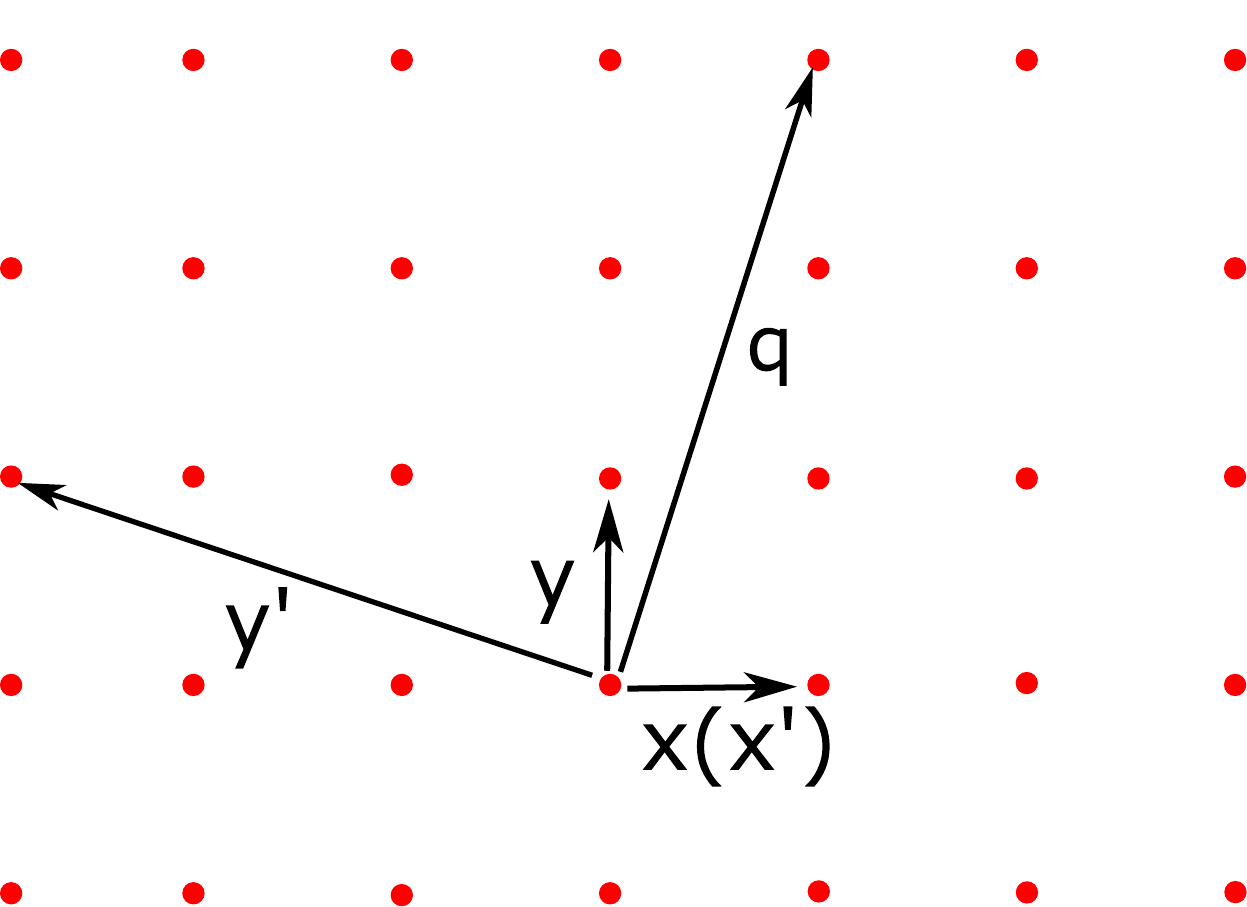}
\caption{ For a DW vector $\bm{q}=|q_{0}|(\frac{1}{\sqrt{1+3^2}},\frac{3}{\sqrt{1+3^2}})$ on the 2D square lattice, we change the basis from $\hat{x}, \hat{y}$ to $\hat{x}', \hat{y}'$ so that we can utilize the translation symmetry along the $\hat{y}'$ direction. }\label{changebasis}
\end{figure}

This wave vector in the 2D DW system corresponds to a magnetic field $\bm{B}=|q_0|(\frac{p}{\sqrt{1+p^2}},-\frac{1}{\sqrt{1+p^2}},0)$ in the 3D system, whose electromagnetic vector potential takes the form of $\bm{A}=(0,0,\bm{q}\cdot \bm{r})$. Consequently, neither $k_{x}$ nor $k_{y}$ is good quantum number. Instead, we can take a new basis:
\begin{equation}
    \begin{split}
        &\hat{x}'=(1,0) \\
        &\hat{y}'=(-\frac{p}{\sqrt{1+p^2}},\frac{1}{\sqrt{1+p^2}}),     
    \end{split}
    \tag{E2}
\end{equation}
so that there remains translation symmetry along the $\hat{y}'$ direction, and $k_{y'}$ is a good quantum number, see Fig.\ref{changebasis} for illustration.

The Hamiltonian $\hat{H}_{2D}=\hat{H}_{0}+\hat{H}_{DW}$ in the new basis takes the same form as Eq. 7 in the main text and can be solved in a similar fashion, with the following changes to the model settings:
\begin{equation}
    \begin{split}
        &\bm{\delta}=\hat{x},\hat{y}\Rightarrow \bm{\delta}=\hat{x}',p\hat{x}'+\hat{y}',\\
        &\bm{\delta}=\hat{x}+\hat{y},\hat{x}-\hat{y}\Rightarrow \bm{\delta}=(p+1)\hat{x}'+\hat{y}',(1-p)\hat{x}'-\hat{y}',\\
        &\bm{q}\cdot\bm{r}\Rightarrow q_{x}x=\frac{|q_{0}|x}{\sqrt{1+p^2}}.\\
    \end{split}
    \tag{E3}
\end{equation}
Finally, we apply a polynomial fit to the resulting $E_0(|\bm{q}|)$ to get rid of the fluctuations in data, mainly caused by the limited system and step sizes, for a smoother display.
\end{document}